\documentclass[%
 reprint,
 amsmath,amssymb,
 aps,
]{revtex4-1}

\usepackage{natbib}
\usepackage{graphicx}
\usepackage{dcolumn}
\usepackage{bm}

\begin{document}

\preprint{APS/123-QED}

\title{Non-local transport via edge-states in InAs/GaSb coupled quantum wells}

\author{Susanne Mueller, Atindra Nath Pal, Matija Karalic, Thomas Tschirky, Christophe Charpentier, Werner Wegscheider, Klaus Ensslin, Thomas Ihn}
\affiliation{Solid State Physics Laboratory, ETH Zurich, 8093 Zurich, Switzerland}

\date{\today}

\begin{abstract}

We have investigated low-temperature electronic transport on InAs/GaSb double quantum wells, a system which promises to be electrically tunable from a normal to a topological insulator. Hall bars of $50\,\mu$m in length down to a few $\mu$m gradually develop a pronounced resistance plateau near charge-neutrality, which comes along with distinct non-local transport along the edges. Plateau resistances are found to be above or below the quantized value expected for helical edge channels. We discuss these results based on the interplay between imperfect edges and residual local bulk conductivity.

\end{abstract}

\pacs{Valid PACS appear here}

\maketitle

After the prediction \cite{BernevigScience2006} and the observation \cite{KoenigScience2007} of the quantum spin Hall (QSH) phase in the two-dimensional topological insulator HgTe/CdTe quantum well (QW) system, there is increased interest in the double QW structure InAs/GaSb sandwiched between AlSb barriers. In this system the overlap between the electron dispersions in the InAs conduction band and in the GaSb valence band is gate tunable \cite{DrndicApplPhysLett1997} due to the spatial separation of the two wells. Hybridization of these two bands at zero magnetic field has been theoretically predicted \cite{AltarelliPhysRevB1983, NavehApplPhysLett1995, LeonPhysRevB1999, MagriPhysRevB2000} and pioneering experiments aimed at verifying this prediction \cite{LakrimiPhysRevLett1997, WagnerSuperlattices1997, YangPhysRevLett1997, KonoPhysRevB1997, CooperPhysRevB1999, YangPhysRevB1999}. More recently a phase diagram was suggested \cite{LiuPRL2008} covering metallic, normal insulator and QSH phases. In the latter counter propagating topologically protected helical edge states are expected to dominate transport properties at zero magnetic field close to the charge-neutrality point (CNP) in devices with reduced structure sizes. This scenario requires edge state scattering to be sufficiently reduced while the bulk is insulating. First studies aiming at the observation of the QSH phase used samples with the highest available material quality \cite{KnezPhysRevB2010, KnezPRL2011, NichelePRL2014, QuArxiv2015}, which still showed substantial residual bulk conductivity. In subsequent devices disorder was intentionally introduced \cite{CharpentierApplPhysLett2013, DuPRL2015} based on the hope of suppressing the bulk conductivity without affecting the topologically protected edge states. In these samples the four-terminal resistance peak at charge-neutrality shrinks when the device dimension is reduced, gradually forming a plateau \cite{KnezPRL2012, SuzukiPRL2013, KnezPRL2014, DuPRL2015}. The expected quantized edge-resistance value of $h/e^{2}$ between neighboring voltage contacts should be reached on samples smaller than the spin-relaxation length of the helical edge states. It is expected to remain insensitive to further reduction of device size as long as edge channels at opposite sample edges do not overlap.

\begin{table*}[t]
\begin{center}
\begin{tabular}{l | l | l | l | l | l| l | l | l | l | l | l}
		Device & $W$ & $L$ & $W_\mathrm{gate}$ & $L_\mathrm{gate}$ & $L_\mathrm{edge}$ & Hysteresis & L-R config. & NL-R config. & NL-R config. & $R_\mathrm{E}$ & $R_\mathrm{C}$\\
		name & [$\mu$m] & [$\mu$m] & [$\mu$m] & [$\mu$m] & [$\mu$m] & $\Delta V$ [V] & [k$\Omega$] & type 1 [k$\Omega$] & type 2 [k$\Omega$] & [k$\Omega$] & [k$\Omega$]\\
   \hline                       
   A & 2.1 & 3.3 & 9.3 & 11.3 & 40.3 & 6.6 & 10.8$\pm$0.5 (84$\%$) & 3.1$\pm$0.3 (72$\%$) & 6.1$\pm$0.4 (71$\%$) & 21.9 & 339.5\\
   B & 2.2 & 5.1 & 9.5 & 13.2 & 41.8 & 4.6 & 11.6$\pm$0.0 (97$\%$) & 2.9$\pm$0.1 (67$\%$) & 5.7$\pm$0.1 (66$\%$) & 23.3 & 193.4\\
   C & 4.4 & 4.4 & 6.5 & 7.4 & 12.9 & 1.0 & 5.4$\pm$0.6 (42$\%$) & 1.5$\pm$0.0 (35$\%$) & 3.0$\pm$0.0 (35$\%$) & 10.7 & 159.8\\
	 D & 3.5 & 5.9 & 6.2 & 9.9 & 19.5 & 0.9 & 7.5$\pm$0.3 (58$\%$)& 1.5$\pm$0.2 (35$\%$)& 2.9$\pm$0.4 (34$\%$) & 15.3 & 65.4\\
   \hline  
   E & 4.9 & 10 & 11.7 & 19.6  & 53.4 & 5.6 & 130.9$\pm$9.1 & 22.7$\pm$2.4 & 45.7$\pm$4.5 & not valid & not valid\\
   \hline  
	 F & 25 & 50 & & & & 0.4 & 90,000 & - & - & - & -\\
   \hline 
\end{tabular}
\caption{Summary of device details as explained in the main text. The meaning of the length scales and the hysteresis $\Delta V$ are indicated in Fig.~1(b) and Fig.~1(e), respectively.}
\label{tab:1}
\end{center}
\end{table*}

Here we report edge-dominated transport where the resistance at charge-neutrality measured between neighboring contacts along the sample edge falls below the expected quantized value even though a bulk resistance of the order of 10 $\mathrm{M\Omega}$ is seen in large area devices. The presence of non-local edge conduction is detected in measurement configurations, where a local conductivity model delivers a voltage drop between contacts too small to be experimentally detectable. Mesoscopic samples show pronounced non-local resistances scaling according to the expectations for helical edge modes without giving the precisely quantized values expected from theory. Based on a resistor network model we separate qualitatively edge and bulk resistances explaining the deviation of the experimentally observed plateau height from the theoretical prediction.

Our devices were fabricated on MBE grown wafers containing an 8\,nm GaSb QW on top of a 15\,nm InAs QW embedded between AlSb barriers. These wafers were grown using a source with reduced Ga-purity as described in Ref.~\onlinecite{CharpentierApplPhysLett2013}. Results from six Hall bars (denoted device A to F, see Table~I) with different lengths $L$ between contacts and widths $W$ (for an exemplary image including device dimension notations see Fig.~1(b)) are discussed in this paper \cite{comment1}. Argon plasma etching was used for the large area devices E and F to fabricate the Hall bar mesa. The smaller devices A to D were patterned by wet etching as reported in Ref. \onlinecite{PalArXiv2015}. All devices were passivated with a 200\,nm thick Si$_{3}$N$_{4}$ dielectric. Gate tunability is implemented with a Ti/Au top gate of width $W_\mathrm{gate}$ and length $L_\mathrm{gate}$ listed in Table~I.

The experiments were performed at a temperature of 1.5\,K and devices A and B were additionally tested at 125\,mK. Reducing the temperature by an order of magnitude resulted in an insignificant enhancement of the longitudinal resistance by less than 5\,$\%$ similar as in Ref. \onlinecite{KnezPRL2014}. Four-terminal resistance measurements were performed by applying a DC current of 5-10\,nA between two contacts and by measuring the DC voltage between two different contacts using a low-noise (30\,nV/$\sqrt{\mathrm{Hz}}$) voltage preamplifier. The resistance of device F, which was highly insulating at charge-neutrality, was measured in a two-terminal voltage-biased configuration with an IV-converter.

\begin{figure}[h!]
  \includegraphics[]{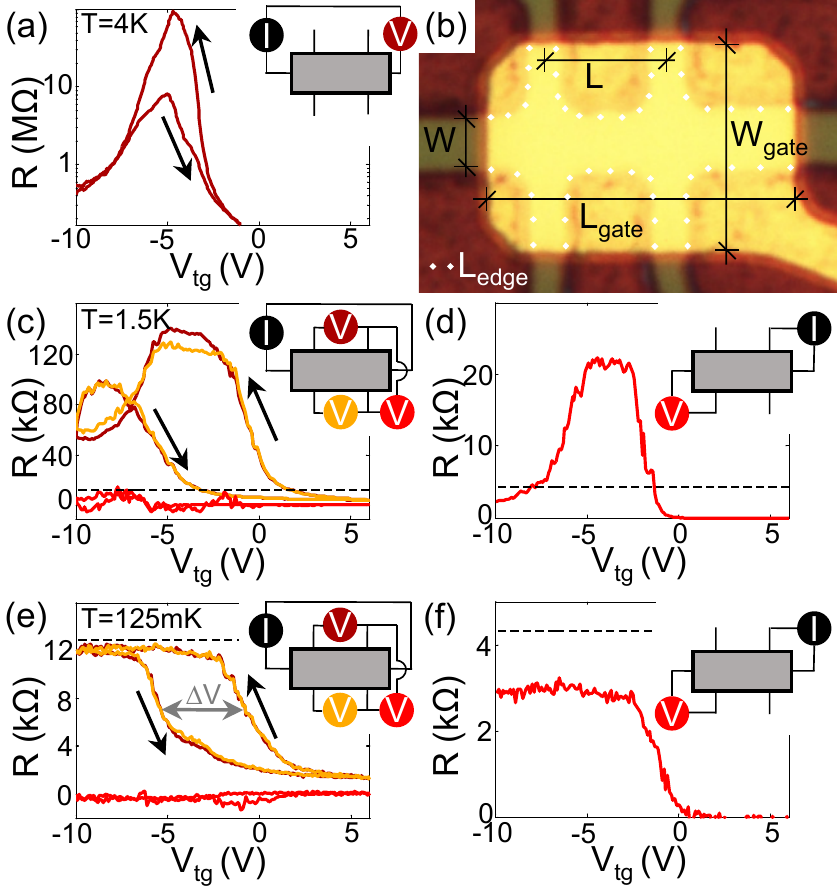}
  \caption{Comparison of two-terminal resistance vs. top gate voltage of a large device ($L=50\,\mu$m, device F) (a) with the four-terminal resistances of a medium sized device ($L=10\,\mu$m, device E) (c) and a small device (typical dimension $L=5\,\mu$m, device B) (e). The resistance peak around the CNP decreases with reduced structure size and forms a plateau. Black arrows indicate the top gate sweep direction. The optical microscope image of device B in (b) indicates the length $L$, width $W$, gate length $L_\mathrm{gate}$, gate width $W_\mathrm{gate}$ and the total length of the gated edge $L_\mathrm{edge}$ used in Table~I. Four-terminal resistances of devices E and B aiming at the measurement of non-local resistances are shown in (d) and (f), respectively. The schematics in each panel indicate how the current source and the voltmeters were connected for the measurements shown. The theoretically expected resistance quantization values of 12.9\,k$\Omega$ in case of the local configuration (a,c,e) or 4.3\,k$\Omega$ for the non-local measurements (d,f) are indicated with horizontal black dashed lines.}
\end{figure}

In all measurements shown in Figs.~1 and 2 we start the top-gate sweep at $V_\mathrm{tg} = +6$\,V, where the resistance $R$ is governed by transport in the conduction band of the InAs QW. Upon lowering the top gate voltage, the resistance increases indicating that the electron density is reduced and the system approaches the CNP. Below $V_\mathrm{tg} = -2$\,V the behavior of the resistance depends strongly on device size as seen in Figs.~1(a,c,e). 

Before we look into these differences in detail, we discuss the hysteresis of the resistance $R$ between down- and up-sweep of the gate voltage shown in Fig.~1(a,c,e) (black arrows indicate sweep direction). The hysteresis is probably due to the accumulation of charge in the gate insulator, or at the insulator-semiconductor interface. The approximate gate voltage shifts $\Delta V$ between up- and down-sweeps are summarized for all devices in Table~I. The gate hysteresis of device B shown in Figs.~1(e) represents one of the worst cases ($\Delta V = 4.6\,$V) whereas in the best case $\Delta V \textless 1\,$V (device D). For large values of gate hysteresis plateau-like features may arise, since the resistance becomes independent of gate voltage. 

Three reasons make us confident that the plateaus observed in Figs.~1(e,f) are still related to the topological properties of the material: First, there is a continuous evolution from the resistance maximum in the largest device F (Fig.~1(a)) via the intermediate device E (Fig.~1(c)) showing a clear plateau-like maximum, to the smallest devices. Second, the observed conductance behavior is well conceivable, if the hole mobility is rather low compared to the electron mobility, and if the contribution of the edge to the conductance scales with edge length. Third, the plateau comes along with a distinct non-local edge conductance, as shown below. Many publications on two-dimensional topological insulators \cite{KoenigScience2007, DuPRL2015, KoenigJPSJ, RothScience2009} show a marginal decrease in resistance for decreasing gate voltage beyond the CNP. Difficulties to tune into the deep hole regime were explicitly reported \cite{KoenigJPSJ}. A good gate tunability in the hole regime is shown in Refs. \onlinecite{RothScience2009, SuzukiPRL2013}. We took great care that the plateaus which we discuss in the following are not related to experimental artifacts by following a consistent measurement protocol. It turned out that measurements are stable and reproducible, if the gate voltage is repeatedly swept in the same range at constant rate. Data presented in this paper are discussed and analyzed only for down-sweeps of the voltage, and the plateaus seen below $V_\mathrm{tg} = -2$\,V are considered to be reliable only down to $V_\mathrm{tg} = -5.5$\,V in case of device B (replication of plateau during up-sweep).

In Fig.~1(a) the two-terminal resistance $R$ in device F as a function of top gate $V_\mathrm{tg}$ shows a profound charge-neutrality resistance-peak of about 90\,M$\Omega$. A continuous transition between the electron and the hole regime with a sign change of the Hall slope in four-terminal magneto-transport measurements (not shown) is observed while crossing the CNP. The peaked resistance transforms into a plateau at 131\,k$\Omega$ for a reduced structure size (device E, shown in Fig.~1(c)), well above the quantization value of $h/2e^{2}$ expected for helical edge states. An even lower and well defined resistance plateau is found in all the small area devices (device A to D, device B is shown exemplarily in Fig.~1(e)). Contrary to the large area device F, a sign change of the Hall slope could not be experimentally detected for the intermediate and smallest samples, an observation whose origin remains to be investigated.

The resistance plateau in Fig.~1(e) stays below the expected quantization value indicated with a dashed black line, reaching about 97\,$\%$ of its anticipated height. The plateau height for the other devices A, C and D are even lower and can be found in Table~I, column ``local resistance configuration'' (L-R config.).  

\begin{figure}
  \includegraphics[]{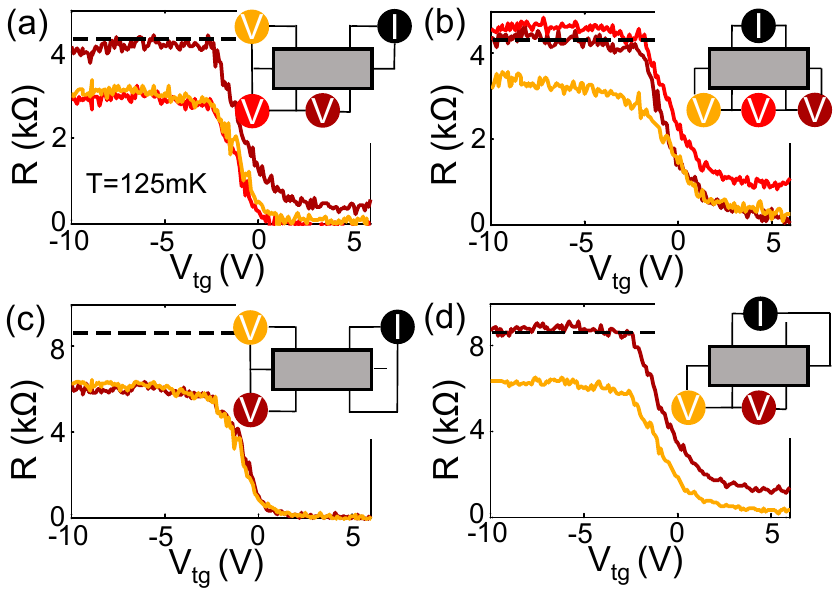}
  \caption{Four-terminal non-local configurations on device B are shown and schematically explained in the respective upper right corners. For current flow between neighboring contacts (a,b) 4.3\,k$\Omega$, respective for a flow between next nearest neighbors (c,d) 8.6\,k$\Omega$ is theoretically expected, indicated with black dashed lines.}
\end{figure}

The predicted transport along helical edge channels is governed by the expectation of a strongly non-local resistance close to charge-neutrality. Non-locality is best probed in four-terminal measurement configurations in which an entirely local resistivity model would give a vanishingly small resistance (see e.g. the inset of Fig.~1(d)). Device E shows a profound non-local plateau of 22.7\,k$\Omega$ when the Fermi energy is tuned to the CNP as shown in Fig.~1(d). Device B (see Fig.~1(f)) shows a reduced plateau height below the expected quantization value of $h/6e^{2}$ (black dashed line) \cite{comment2}.

For a detailed non-locality discussion we concentrate now on device B and the data shown in Fig.~2. In Fig.~2(a) an example of a measured vanishing local response at $V_\mathrm{tg} > 0.5$\,V (red and yellow traces) is shown, where the Fermi energy is deep in the conduction band and transport is well described by a local resistivity model. Comparing the red and yellow traces to the brown trace in the same voltage range, we realize that the measured four-terminal resistance grows when the pair of voltage probes is closer to the current carrying contacts. This corresponds to the fact that the current density flowing through the bulk of the sample weakens strongly with increasing distance from the current contacts. Note also that within a homogeneous local resistivity model, the bulk current density distribution does not depend on the value of the resistivity.

When the gate voltage is tuned close to the CNP ($V_\mathrm{tg} < -1$\,V) in Fig.~2(a), the measured resistance shows a plateau. The red and yellow traces saturate at the same plateau value, whereas the brown trace shows its plateau at a larger resistance. In the picture of edge transport, the same current flows along the sample edges between all four involved voltage contacts, and we would therefore expect the same plateau resistance to be measured in all three configurations. The enhanced plateau resistance of the brown curve is in qualitative agreement with the notion of a higher bulk contribution to the current leading together with the edge current to a larger voltage drop along the sample edge. Concomitantly a larger length of the physical edge may lead to a larger edge-resistance. We draw the preliminary conclusion, that in our samples there is strong evidence for non-local transport along the edge, which is, however, still influenced by a finite residual bulk conductivity having a local character. We may call the red and yellow traces truly non-local on the plateau, because the contribution of the local bulk conductivity vanishes for the corresponding voltage contacts, as confirmed by the zero resistance value deep in the electron regime.

This picture has to be stated more precisely, if we compare the plateau values of the three traces in Fig.~2(a) quantitatively to the theoretical expectation for transport in ideal helical edge channels (black dashed line). The experimental fact that the red and yellow traces show a plateau at 67\,$\%$ of the expected value seems to indicate that the current in the corresponding edge segments is reduced below its ideal value, probably because the finite bulk conductivity diverts some current away from these edge segments. For the brown trace, however, which results from an edge segment closer to the current contacts, it seems that either the additional bulk current just about compensates for this reduced edge current leading to a close to ideal plateau value, or that the edge-resistance is larger due the larger length of the edge. Within this picture the ideal plateau value appears rather like an accidental coincidence than a cogent effect.

We have performed similar four-terminal measurements of all possible contact combinations of current and voltage leads. In general we find that different contact configurations related by the generalized Onsager symmetry relations \cite{Buttiker1986} always give consistent results, as expected. This reduces the set of independent four-terminal measurements in our six-terminal devices to ten, which classify into conventional configurations (Fig.~1(e)), non-local resistance configurations of type 1 (Figs.~2(a,b)), where the current is driven between neighboring contacts, and non-local resistance configurations of type 2 (Figs.~2(c,d)), where the current flows between next nearest neighboring contacts.

Additionally, geometric symmetries of the samples (e.g. reflections at the Hall bar axis, or inversions at the Hall bar center) always gave consistent results. This observation and the vanishing zero magnetic field Hall resistance (see Figs.~1(c,e)) give evidence for a homogeneous bulk and excellent contact properties witnessing the high quality of our devices. 

Due to the Onsager symmetries the sketched schematics in Fig.~1(e) and Figs.~2(a-c), represent all the possible configurations. In Fig.~1(e) and Fig.~2 the expected quantization values for helical edge states of $h/2e^2$ (Fig.~1(e)), $h/6e^{2}$ (Figs.~2(a,b)) and $h/3e^{2}$ (Figs.~2(c,d)) are shown as dashed black lines. In all configurations the experimental plateau values are systematically below the expectation. For all voltage drops, which are truly non-local (far away from the current leads, vanishing local resistance in the electron regime, see Fig.~2(a) red and yellow, and Fig.~2(c)), the ratios of plateau values of pairs of different configurations are exactly given by the ratios of the expected plateau values in an ideal topological insulator without residual bulk conductivity. Device B, for example, reaches about 66\,$\%$ of the expected plateau values in all truly non-local configurations. The mean plateau values together with their uncertainties and these scaling factors are summarized in Table~I for all devices and all types of configurations. The clear non-local signals together with this consistent scaling are in agreement with the theoretically proposed current carrying helical states along the edge.

\begin{figure}
  \includegraphics[]{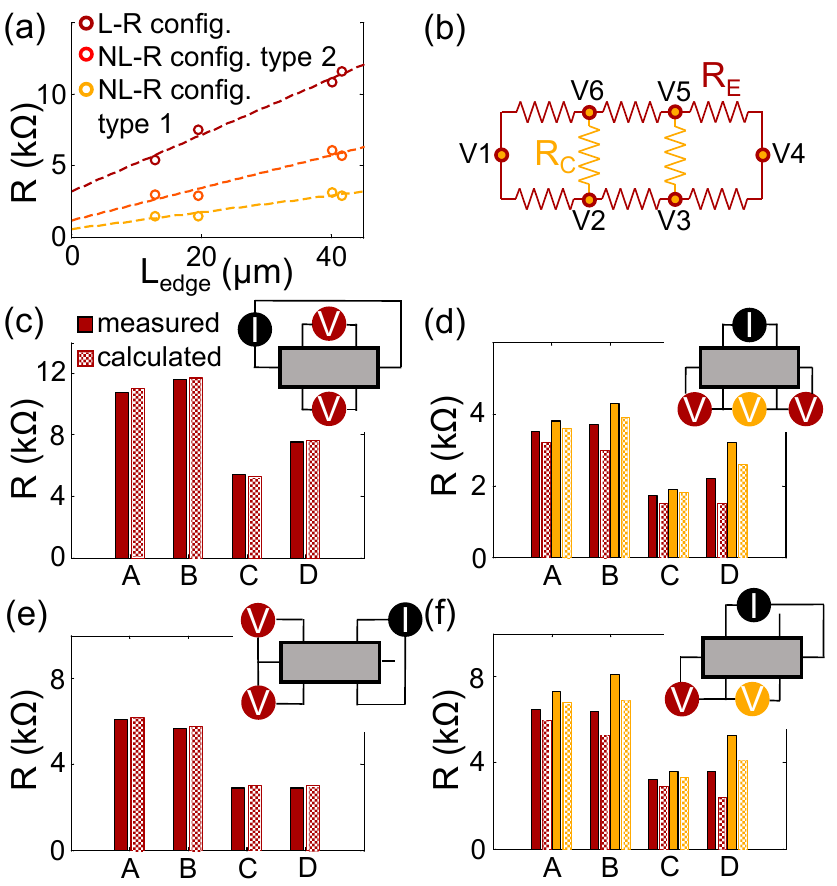}
  \caption{The linear dependence of the local (L-R) as well as non-local (NL-R) four-terminal resistances with the gated edge length $L_\mathrm{edge}$ (a) suggests the resistor network illustrated in (b). Here $R_\mathrm{C}$ symbolizes a residual bulk resistance, $R_\mathrm{E}$ the edge-resistance and $\mathrm{V_1}$ through $\mathrm{V_6}$ the voltage probes. The bar graphs in (c)-(f) allow a comparison of samples A through D between the measured four-terminal resistances (filled bars), as schematically shown in the respective top right corners, and the calculated prediction with the model (dotted bars).}
\end{figure}

Figure~3(a) shows that the plateau value of the longitudinal resistance in the conventional measurement configuration increases linearly with the edge length. The same trend can be seen for the truly non-local resistances of the configurations of type 1 and 2. This finding suggests a resistor network model for the description of transport in the plateau region of the devices consisting of a series of resistors $R_\mathrm{E}$ along the edge as schematically shown in Fig.~3(b) in red. The observed finite bulk conductance is accounted for by adding bulk leakage resistors $R_\mathrm{C}$ to the model in Fig.~3(b). Although this model oversimplifies the real situation it may serve as a tool to compare the relative relevance of edge and bulk conduction.

One measurement configuration (here the configuration shown in Fig.~2(a)) is sufficient to calculate the two unknown resistors in the network model. The result is a bulk coupling $R_\mathrm{C}$ between 65.4\,k$\Omega$ (device D) and 339.5\,k$\Omega$ (device A), which is an order of magnitude larger than the edge-resistance $R_\mathrm{E}$ ranging from 15.3\,k$\Omega$ (device D) to 21.9\,k$\Omega$ (device A) (for more details see Table~I), confirming the dominant edge conduction and the model assumption. Even though, $R_\mathrm{C}$ simplifies the description of the bulk contribution, the calculated predictions of all possible measurement configurations based on the configuration in Fig.~2(a) pass the test of being compared with the measurements as illustrated with the bar graphs in Figs.~3(c-f). Our analysis demonstrates that one set of $R_\mathrm{E}$ and $R_\mathrm{C}$ is sufficient to describe all measurement configurations. Since $R_\mathrm{E}$ is consistently below the expected $h/e^{2}$, it seems likely that $R_\mathrm{E}$ is a combination of bulk conduction and the edge conduction of $h/e^{2}$. 

To conclude, we have investigated the edge conduction in standard Hall bar samples of InAs/GaSb QWs. At the CNP a pronounced non-local resistance below the expected quantization value was observed in all devices with structure size below 6\,$\mu$m together with a consistent scaling according to Landauer-B\"uttiker theory, suggesting edge conduction with helical character. The deviation from the theoretical prediction could be described qualitatively in a resistor network model.

The authors acknowledge helpful discussions with L. Glazman and C. Lewenkopf, technical support by P. M\"arki and thank the Swiss National Science Foundation for financial support via NCCR QSIT (Quantum Science and Technology).


\begin{thebibliography}{99}
\bibitem{comment1} Analogous experiments (not shown) as presented in this manuscript but performed on a sample of similar dimensions fabricated on a low-disorder wafer did neither display non-local transport effects nor plateau-like features in the charge-neutrality regime.
\bibitem{comment2} Since measurements on the high-resistance device F are rather cumbersome, so far no convincing non-local data could be detected.
\end{thebibliography}
\end{document}